\documentstyle[aps, psfig]{revtex}
\newcommand{\be}{\begin{equation}}
\newcommand{\ee}{\end{equation}}
\newcommand{\bea}{\begin{eqnarray}}
\newcommand{\eea}{\end{eqnarray}}
\newcommand{\bm}{\bibitem}

\newcommand{\al}{\alpha}
\newcommand{\bet}{\beta}
\newcommand{\gm}{\gamma}
\newcommand{\gz}{\gamma_0}
\newcommand{\Gm}{\Gamma}
\newcommand{\gf}{\gamma_5}
\newcommand{\ep}{\epsilon}
\newcommand{\gmd}{\gamma_{\mu}}
\newcommand{\gmu}{\gamma^{\mu}}
\newcommand{\de}{\delta}
\newcommand{\om}{\omega}
\newcommand{\omk}{\omega_k}
\newcommand{\omp}{\omega_p}
\newcommand{\th}{\theta}
\newcommand{\Th}{\Theta}
\newcommand{\tpp}{\theta (p_0)}
\newcommand{\tmp}{\theta (-p_0)}
\newcommand{\lm}{\lambda}
\newcommand{\Lm}{\Lambda}
\newcommand{\sg}{\sigma}
\newcommand{\Sg}{\Sigma}
\newcommand{\bq}{\bar{q}}
\newcommand{\bu}{\bar{u}}
\newcommand{\bd}{\bar{d}}

\newcommand{\bn}{\overline{n}}
\newcommand{\hm}{\hat{m}}
\newcommand{\us}{u \!\!\! /}
\newcommand{\xs}{x \!\!\! /}
\newcommand{\ps}{p \!\!\! /}
\newcommand{\qs}{q \!\!\! /}

\newcommand{\bt}{{\cal B}_M}    
\newcommand{\vp}{\vec{p}}
\newcommand{\vq}{\vec{q}} 
\newcommand{\vk}{\vec{k}}
\newcommand{\la}{\langle}
\newcommand{\ra}{\rangle}
\newcommand{\rw}{\rightarrow}

\newcommand{\F}{F^2_{\pi}}

\begin{document}

\setcounter{page}{1}

\title{QCD sum rules for nucleon two-point function in nuclear medium}

\author{S. Mallik} 
\address{Saha Institute of Nuclear Physics,
1/AF, Bidhannagar, Kolkata-700 064, India} 

\author{Hiranmaya Mishra} 
\address{Theory Divison, Physical Research Laboratory, Ahmedabad 380 009, 
 India}

\date{\today} 

\maketitle

\begin{abstract} 

We derive QCD sum rules from the nucleon two-point function in nuclear
medium, calculating its specral function in chiral perturbation theory
to one loop. Our calculation shows the inadequacy of the commonly used
ansatz to represent the two-point function as the sum of a nucleon pole 
term and a spectral integral over the high energy continuum. We also point
out that it is the energy variable and not its square that must be used in
writing the dispersion integrals. The sum rules are obtained by equating 
terms linear in nucleon number density of the spectral and operator
representations of the correlation function. We use them to predict the
curent-nucleon coupling, in good agreement with the value known from the
vacuum sum rules. They also predict a definite deviation of the four-quark 
condensate from its ground state saturation value.

\end{abstract}


\section{introduction}

The extension of the original QCD sum rules \cite{SVZ} to finite temperature
and chemical potential has been the subject of many investigations 
\cite{Hatsuda,Drukarev1}.  These
were first extended to finite temperature by Bochkarev and Shaposhnikov
\cite{Shap}, who pointed out the importance of continuum contribution in the
low energy region. Working with the two-point function of vector currents, they
showed that the contribution of the $\pi\pi$ intermediate state, though
negligible in vacuum compared to that of the $\rho$-pole, is important at
finite temperature.

As shown by Leutwyler and Smilga \cite{Leutwyler90}, chiral perturbation
theory provides a reliable method to calculate the correlation functions.
They took the two-point function of nucleon currents at finite temperature
and considered all the one-loop Feynman graphs of this
theory to calculate the nucleon pole parameters at finite temperature. Later
Koike \cite{Koike} worked out directly the absorptive parts of these graphs 
to construct their spectral representations. He then evaluated the QCD sum 
rules in agreement with known results. 

The sum rules at finite temperature were subsequently taken up in Refs.
\cite{Mallik1,Mallik2}, where their spectral sides were calculated from 
the set of all one-loop Feynman graphs with vertices from chiral 
perturbation theory. It was shown that the contribution 
of each graph can, in general, be separated into a pole term for the 
communicating single particle and a remainder, that is finite at the pole. 
Clearly to evaluate the changes in the pole paramters, the remainder terms 
are not necessary. But in the evaluation of the sum rules, where we require 
their contributions at large external momenta, both the pole term and the 
remainder are of comparable magnitude.

This observation questions the widely used ansatz for the two-point function
given by a pole term (with modified parameters) together with continuum
contributions from {\it high} energy region only. Such an Ansatz ignores the
equally important contribution from the contiuuum in the {\it low} energy 
region and is therefore incomplete.

In this work we investigate the QCD sum rules derived from the two-point
function of nucleon currents at finite nucleon chemical potential 
\cite{Drukarev2,Jin}. Our work concerns the spectral side of the sum rules. 
As at finite temperature, we work out the spectral representation in  
nuclear medium in chiral perturbation theory to one loop. The contribution 
of the Feynman graphs are separared into the nucleon pole term and a 
continuum contribution in the low energy region. Again we find the 
Borel transform of these two types of terms to be of comparable magnitude.

Another point overlooked in the literature is the absence of any symmetry
of the amplitudes in the energy variable in the present case. Unlike in 
pionic medium (at finite temperature), the two-point function in nuclear 
medium has no density dependent singularities from the `crossed channel', 
as we have no antinucleons in the medium. Thus if $q$ is the external 
four-momentum and we restrict to $q_{\mu}=(q_0=E, \vq=0)$, there is no 
absorptive part of the amplitudes for $E<0$. Clearly the amplitudes have 
no symmetry under $E\rw -E$. As a result the dispersion variable must be 
$E$ and not $E^2$.

We also address the issue of sensitivity of the sum rules. The in-medium sum
rules, as they are generally written, are actually the corresponding vacuum
sum rules, to which are added density dependent contributions. If the
vacuun sum rule is stable in a region of the Borel parameter and the density
dependent `corrections' are small, as is usually the case, it is likely that
the stability of the in-medium sum rules do not depend sensitively on 
these corrections.

To restore the sensitivity of our sum rules, we equate 
terms linear in the nucleon number density on the spectral and
the operator sides of the two-point function. Our in-medium sum rules are
thus independent of the vacuum sum rule, being built out of the `correction 
terms' only. We recall here that in the vacuum sum rules the unit operator 
provides the dominant contribution on the operator side, to which higher 
dimension operators add small `power corrections'. In the present sum rules, 
the contribution of the unit operator, being independent of density, is 
absent altogether.

In Sec.II we discuss the ingredients needed to write our sum rules.  Here we
present the form of the spectral representation given by the real-time field
theory in medium. Then we derive the interaction vertices from
chiral perturbation theory. We also list here the operators and their Wilson
coefficients, that we shall include in expanding the operator product in the
two-point function. The spectral representations of different Feynman graphs
are worked out in Sec.III. The sum rules are written and evaluated in Sec.IV. 
In Sec.V we comment on different aspects of our sum rules. 
 
\section{preliminaries}
\setcounter{equation}{0}
\renewcommand{\theequation}{2.\arabic{equation}} 

\subsection{Two point function}

Although we restrict the sum rules to symmetric  nuclear matter at zero 
temperature, our procedure to derive these is completely general. Thus we 
start with the ensemble average of the time ordered product of two nucleon 
`currents',
\be
\Pi (q)=i\int d^4x e^{iqx} \la T\eta (x) \bar{\eta} (0)\ra ,
\ee
where for any operator $O$, $\la O\ra= Tr\left[ \rho O \right]/Tr \rho~, 
\rho = e^{-\beta(H-\mu N)}$, $H$ being the Hamiltonian of 
the system, $\beta$ the inverse of temperature and $N$ the nucleon number 
operator with chemical potential $\mu$. The nucleon 
current $\eta (x)_{D,i}$, with spin and isospin indices $D,\,i$ respectively, 
is built out of three quark fields, so as to have the quantum numbers of the 
nucleon \cite{Ioffe1}. In obtaining the spectral representation, we 
need only its hadronic matrix elements, but to expand the operator product 
in (2.1) into local operators, we must know its specific structure. Of 
the different possibilities, we take here the preferred one \cite{Ioffe2}, which 
for proton ($i=1$) is 
\be
\eta_{D,1}(x)=\epsilon^{abc}\left(u^{aT}(x)C\gmu
u^b(x)\right)\left(\gf\gmd d^c(x)\right)_D\,,
\ee
where $C$ is the charge conjugation matrix and  $a,\, b,\,c$ are the color 
indices.

It is well-known that in a medium amplitudes take a Lorentz covariant form, 
if we introduce the four-vector $u^\mu$, the four-velocity of 
the medium. Thus our two point function admits a Dirac decomposition,
$\Pi(q)=A(q){\bf 1} +B(q)\qs+C(q)\us$. In what follows we write the sum 
rules for $\vec q=0$, when it reduces to $\Pi(q_0=E,\vec q=0)=A(E){\bf 1}
+D(E)\gz$.

We shall use the real time version of the field theory in a medium, where 
a two point function assumes the form of a 2$\times$2 matrix. But the dynamics 
is given essentially by a single analytic function, that is determined by 
the $11$-component itself. Thus if $F_{11}(q)$ is the $11$-component of the
matrix amplitude, the corresponding analytic function $F(q)$ has the spectral 
representation \cite{Landsmann}
\be
F(E,\vec q)=\frac{1}{\pi}\int \frac{\coth(\bet (E-\mu)/2) Im F_{11}(E',\vec q)}
{E'-E-iE'\ep} dE'.
\ee

As the chemical potential for particles and antiparticles are equal and
opposite, its presence in Eq.(2.1) leads to a loss of symmetry of the amplitude 
under $E\rw -E$. The simplest example to observe this asymmetry is provided
by the components of the $2\times 2$ matrix of the free nucleon propagator.  
Denoting the nucleon field by $\psi(x)$, we have for the $11$-component
\cite{Kobes},
\be
\frac{1}{i} S(p)_{11} \equiv \int d^4x e^{ipx}\la T\psi(x)\bar\psi(0)\ra_{11}
=(\ps+m)\left[\frac{i}{p^2-m^2+i\ep}-\left\{n^-(\omp)\tpp+
n^+(\omp)\tmp\right\}2\pi\delta(p^2-m^2)\right],
\ee
where $n^{\mp}$ are the distribution functions for nucleons and
antinucleons, $ n^\mp(\omp)=\left\{e^{\bet(\omp\mp\mu)}+1\right\}^{-1},\, 
\omp=\sqrt{m^2+\vp\,^2}$, $m$ being the nucleon mass. 
The asymmetry is particularly explicit in the limit $\beta^{-1}\rw 0$, 
in which we shall work, when $n^-(\omp)\rw \theta(\mu-\omp)$ and 
$n^+(\omp)\rw 0$. 
As we shall see in the next section, for the two-point function (2.1) at
$\vq =0$, this has the consequence that the analytic function (2.3) has 
no imaginary part on the negative real axis in the $E$ plane. The range 
of integration therefore extends only over 0 to $\infty$.

Note that the analytic function $S(E)$ corresponding to the free 
propagator-matrix in medium is independent of $n^{\mp}$, coinciding with 
the free propagator in vacuum. But in keeping with the absence of 
singularities for $E<0$ from loop diagrams, we replace it by (for $\vq =0$)
\be
S(E) \rw -\frac{\gz E +m}{2m} \frac{1}{E-m+i\ep}.
\ee

\subsection{Chiral perturbation theory}
To calculate the two point function (2.1), it is most convenient to use 
the external field method, where the nucleon current $\eta(x)$ is coupled 
to an external spinor field $f(x)$. The well-known QCD Lagrangian 
${\cal L}_{QCD}^0$ is thus extended to
\be
{\cal L}={\cal L}_{QCD}^0+\bar f\eta+\bar\eta f.
\ee

Chiral perturbation theory \cite{Gasser1} gives the effective Lagrangian 
for the pion nucleon system in terms of the pion matrix field 
$U(x) \in  SU(2)$ and the Dirac spinor $\psi(x)$ for the nucleon 
\cite{Gasser2}. In what follows we ignore isospin symmetry
breaking. At leading order, the effective Lagrangian in presence of the external 
field $f(x)$ is given by,
\be
{\cal L}_{e\!f\!f}= {\cal L}_\pi + \bar\psi(i {\partial \!\!\! /} -m)\psi
+\frac{1}{2}g_A\bar\psi\us\gf\psi+{\cal L}_f~,
\ee
where ${\cal L}_\pi$ is the well-known Lagrangian for pions, whose
interactions we do not need in this work. The matrix field $u$ is given 
by \footnote{No confusion should arise from using the same $u$ to denote the
pion matrix field (only in this subsection) and the $u$-quark field as well as
the four-velocity $u^{\mu}$.},
$u^2=U$ and $g_A$ is the axial vector constant ($g_A=1.27$). ${\cal L}_f$ 
is the part involving the external field $f(x)$. Taking note of the 
transformation properties of different fields, we can write it to lowest order as
\be
{\cal L}_f=\lambda\bar f(u\psi_L+u^\dagger \psi_R)+c.c\,,
\ee
where $\psi_{R,L}$ are the right- and left-handed components of $\psi$, defined 
as $\psi_{R,L}=\frac{1}{2}(1\pm\gf)\psi$ and $\lm$ is the current-nucleon 
coupling constant.

In the explicit representation $U=\exp(i\phi^a\tau^a/F_\pi)$, where
$\phi^a$ are the hermitian pion fields ($a=1,2,3$), $\tau^a$ the Pauli 
matrices and $F_\pi$ the pion decay constant ($F_\pi=93$ MeV),  we have
\be
{\cal L}_{e\!f\!f}=\bar\psi(i{\partial \!\!\! /}-m)\psi-
\frac{g_A}{2F_\pi}\bar\psi\gm_\mu\gf\tau^a\psi\partial^\mu\phi^a
+\lm\bar f\left(\psi-\frac{i\tau^a}{2F_\pi}\gf\psi\phi^a\right) + c.c.
\label{leff}
\ee

We need two more vertices. One is the four-nucleon coupling derived by 
Weinberg \cite{Weinberg}. Written in terms of the four component Dirac 
spinor $\psi$, this nonrelativistic Lagrangian has the form
\be
{\cal L}_{\psi^4}=-\frac{C_S}{8}\left[\bar\psi(1+\gm_0)\psi\right]^2
-\frac{C_T}{8}\left[\bar\psi(1+\gz)\vec\gm\gf\psi\right]^2,
\ee
where $C_S$ and $C_T$ may be obtained from the s-wave scattering 
lengths as \footnote{ Here we take the singlet and triplet scattering lengths 
as $-16.4$ fm and $5.40$ fm respectively. These are obtained from $np$ and
$pp$ scattering experiments \cite{Haddock}. Theoretical calculations
based on chiral Lagrangians also reproduce these values \cite{Ordonez}.
But a different value for the singlet scattering length, namely $-23.7$ fm
also exists in the literature \cite{Bohr}. Our sum rules give less
satisfactory results with this latter value.} 
\[C_S = -\left(\frac{1}{506 MeV}\right)^2,~~~~~ 
C_T=\left(\frac{1}{52MeV}\right)^2 .\] 

The other vertex is that of $\eta\rightarrow \psi\bar\psi\psi$. The 
most general form of this coupling would give $\eta$ as
\be
\eta=\sum_i \{a_i(\bar\psi\Gm^i\psi)\Gm_i\psi +
b_i(\bar\psi\Gm^i\tau^a\psi)\Gm_i\tau^a\psi\}~,
\ee
where the sum runs over all the independent Dirac matrices $\Gm^i$. For the
particular Feynman graph we shall be interested in, only one linear
combination of the ten coupling constants $a_i$ and $b_i$ will appear in its
evaluation.

\subsection{Operator product expansion}
As is well-known, not only the Lorentz scalar operators but also the vector and 
tensor operators can have non-zero ensemble averages due to the
availability of the velocity four-vector $u_\mu$ of the medium. Examples are the 
quark current operator $\bar q\gm_\mu q=\bar u\gm_\mu u+\bar d\gm_\mu d$ and 
the (traceless) energy momentum tensor operators $\Th_{\mu\nu}^f$ and 
$\Th_{\mu\nu}^g$ of quarks and gluons respectively given by
\be
\Th_{\mu\nu}^f=\bq i\gm_\mu D_\nu q-\frac{\hm}{4}g_{\mu\nu}\bq q \,,
\ee
\be
\Th_{\mu\nu}^g=-G^c_{\mu\lm}{G_\nu^\lm}^c + \frac{1}{4}g_{\mu\nu}
G_{\al\bet}^c G^{\al\bet c},
\ee
with the symbols having their usual meaning. ($\hm$  is the quark mass in the 
$SU(2)$ symmetry limit.) But as we already mentioned, the unit operator, 
that contributes dominantly to the vacuum sum rule, will not contribute at all 
to our in-medium sum rules.

Thus the contributing operators of lowest dimension in the expansion of the
operator product of nucleon currents are $\bq q$, $\bq \us q$ (=$q^\dagger q$ 
in the rest frame of matter, $u^\mu=(1,0,0,0)$). Then there are the operators 
of dimension four, namely $(\al_s/\pi)G_{\mu\nu}^aG^{\mu\nu a}\,,~~ \Th^{f,g}
\equiv u_\mu u_\nu\Th_{\mu\nu}^{f,g}$. Here we retain only the operator
$\Th^f$, as the coefficients of the remaining operators are 
small, arising from three contractions of the the quark operators in the 
two-point function. Among dimension five operators, the coefficient of
 $\bar q\sigma_{\mu\nu}G^{\mu\nu}q$ turns out to be zero \cite{Mallik1}.
The other operators bring small contributions to the sum rules \cite{Jin}
and are ignored.

Of dimension six operators, we retain, as usual, only the quark operators that 
have no derivatives or gluon fields. In the case of vacuum sum rules, their 
vacuum expectation values may be related to the square of the vacuum expectation 
value of $\bq q$, using the so-called factorisation or vacuum saturation. A 
similar approximation can also be made for the ensemble average, but simple 
model estimates seem to suggest that it may not be as good as in the case of 
vacuum. At this point we follow Ref.\cite{Jin} to use factorisation and then replace 
the scalar-scalar condensate by the parametrisation,
\be 
\la\bq q\ra^2\rw (1-f)\la 0|\bq q|0\ra ^2+ f\la\bq q\ra^2\,,
\ee 
where $ f$ is a real patrameter.

The coefficient of operator product expansion at short distance are evaluated 
conveniently in configuration space. Such a method has been described in 
the Appendix of Ref.
\cite{Mallik1} for the case at hand. One first performs a Wick expansion of the 
two-point quark composite operators. The coefficients of the above mentioned 
operators are obtained from the single and double contractions of the quark fields.  
We state here the results of such a calculation,
\bea
\la T\eta(x)\bar\eta(0)\ra &\rw &\frac{2}{\pi^4x^6}\left\{\la \bu u\ra +
2\left(\us+\frac{2 x.u\xs}{x^2}\right)\la \bar u\us u\ra\right\}\nonumber\\
& - &\frac{4i}{3\pi^4 x^6}\left\{\left(-3+\frac{8(x.u)^2}{x^2}\right) \xs 
+4 x.u \us\right\}\la \Th^f\ra \nonumber\\
&+& \frac{i}{3\pi^2x^4}\left\{\xs\la\bu u\ra^2+2 x.u\la\bu u\ra\la\bu\us
u\ra\right\},
\eea
where we factorize the four-quark operators and omit the piece
$\la \bu \us u\ra^2$, being quadratic in nucleon density. Also we shall ignore
the small effects of anomalous dimensions of the operators and assume their
renormalization scale to be $ 1$ GeV.

\section{spectral representation}
\setcounter{equation}{0}
\renewcommand{\theequation}{3.\arabic{equation}}

Here we want to calculate the (density dependent part of the) two-point 
function in the low energy region to first order in $\bn$, the nucleon
number density in symmetric nuclear matter at zero temperature,
\[ \bn =4\int \frac{d^3p}{(2\pi)^3} \theta (\mu-\omp) =
\frac{2p_F^3}{3\pi^2},\]
where $p_F$ is the Fermi momentum related to the chemical potential $\mu$ by
$\mu^2=m^2+p_F^2$. To this end we draw all the Feynman graphs with one loop 
containing a nucleon line. In addition to the single nucleon line forming a 
loop, we include also the loop containing an additional pion line to account 
for singularities with the lowest threshold. They are depicted in Fig. 1  
along with the free propagator graph (a).

\begin{figure}
\centerline{\psfig{figure=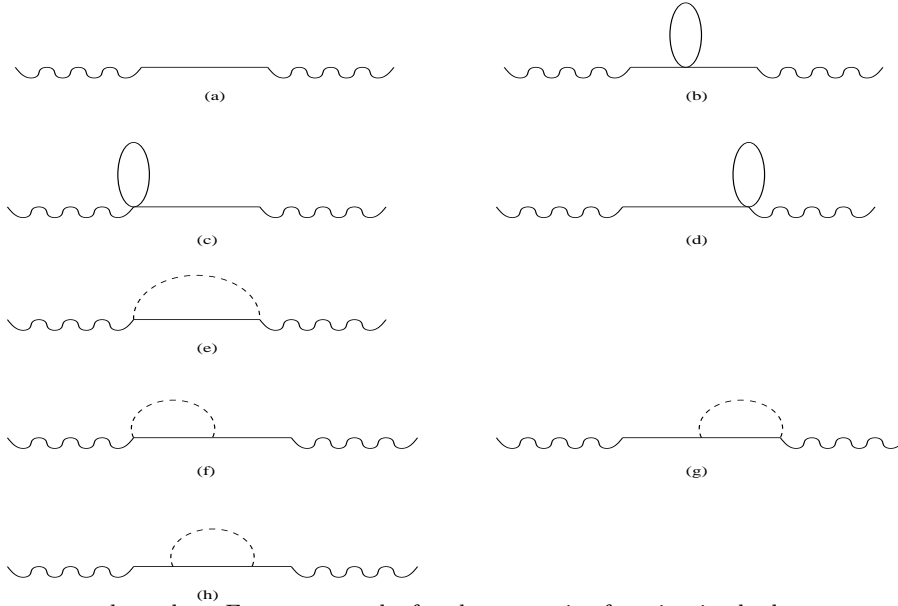,width=12cm,height=8cm}}
\caption{Free propagator and one loop Feynman graphs for the two point 
function in the low energy region. The wavy line represents the nucleon 
current, while the continuous and the dashed internal lines are for nucleon
and pion propagation.}
\end{figure}

Any nonsingular term in the contribution of graphs is of no interest to us
and will be omitted. Thus the free amplitude of graph (a) is given by
\be
\Pi(E)_{(a)}= \lm^2 S(E) = -\frac{\lm^2}{E - m + i\ep}\,\frac{1}{2} (1+\gz),
\ee
on making the replacement (2.5).

To evaluate the self-energy diagram (b) we rewrite the four-nucleon 
interaction term given by Eq. (2.10) conveniently as
\be
{\cal L}_{\psi^4}=-\frac{1}{8}\sum_{i=S,T}C_i\Gm^i_{AB}\Gm^i_{CD}
(\bar\psi_A\psi_B)(\bar\psi_C\psi_D),
\ee
where $\Gm^S=(1+\gm_0)$, and, $\Gm^T=(1+\gz)\vec\gm\gf$.
Then the two point function for this diagram is given by
\be
\Pi(E)_{(b)}=-\lm^2 S(E) \sg S(E).
\ee
Here the self-energy $\sg$ is a constant given by
\be
\sg=-\frac{i}{4}\sum_{i=S,T}C_i\int\frac{d^4p}{(2\pi)^4} 
\{-2tr(S(p)_{11}\Gm^i)\Gm^i+\Gm^iS(p)_{11}\Gm^i\},
\ee
where the $tr$(ace) is over $\gm$-matrices. On inserting the density 
dependent part of $S_{11}$ from Eq. (2.4) it gives
\be
\sg=\frac{3}{8}(C_S-C_T)\bar n (1+\gm_0),
\ee
reproducing the result obtained earlier\cite{Montano}. Eq.(3.3) then gives
\be
\Pi(E)_{(b)}=-\frac{3\lm^2}{8m}(C_S-C_T)\bn\left\{\frac{1}{E-m} +
\frac{m}{(E-m)^2}\right\}(1+\gz).
\ee

It is simple to evaluate the diagrams (c) and (d). Since two of the nucleon
fields are contracted at the vertex itself, we may do so in Eq.(2.11) for $\eta$ 
itself and retain the $\bn$ dependent term to get
\[\eta(x)\rw -(\al+\bet\gz)\bn\psi(x), \]
where $\al$ and $\bet$ depend linearly on $a_i$ and $b_i$. Then we get
\bea
\Pi(E)_{(c)+(d)} & =& -2\lm \bn (\al+\bet \gz) S(E)
\nonumber \\
&=& \frac{\lm^2\zeta\, \bn}{E-m}(1+\gz), ~~~~ \zeta = (\al+\bet)/\lm.
\eea
Here the coupling constant $\zeta$ is unknown. We shall attempt in Sec.IV to
find it from our sum rules.

We now consider the diagram (e) with the $\pi N$ two-particle intermediate 
state. It is the prototype of all the remaining diagrams, so we derive its 
integral representation in some detail. Its contribution is
\be
\Pi(q)_{(e)}=-\frac{3\lm^2}{4\F}\Gm(q)
\ee
where
\be
\Gm(q)_{11}=i\int d^4 x e^{iq.x} \frac{1}{i} D(x)_{11}\gf \frac{1}{i}
S(x)_{11}\gf.
\ee
The standard way  to find the imaginary part of such a loop diagram is to write
it in momentum space and integrate out the time component of the loop momentum, 
from which the imaginary part may be read off. But here this integration is 
difficult due to the presence of $\theta(\pm p_0)$ in the nucleon propagator
(2.4). We may, however, integrate out the $p_0$ variable in the propagator 
itself, getting
\bea
\frac{1}{i}S(x)_{11}&=&\int\frac{d^3p}{(2\pi)^3
2\omp}\Big[\left\{(\ps+m)(1-n^-)e^{-ip.x}
+(\ps-m)n^+e^{ip.x}\right\}\theta(x_0) \nonumber \\
&-& \left\{(\ps+m)n^-e^{-ip.x}+(\ps-m)(1-n^+)e^{ip.x}\right\}\theta(-x_0)\Big].
\eea
The analogous expression for the pion propagator is
\be
\frac{1}{i} D(x)_{11}=\int\frac{d^3k}{(2\pi)^32\omk}\Big[\left\{(1+n)e^{-ik.x}+
n e^{ik.x}\right\}\theta(x_0)
+\left\{ne^{-ik.x}+(1+n) e^{ik.x}\right\}\theta(-x_0)\Big],
\ee
where $n$ is the pion distribution function at temperature $\bet^{-1}$, 
$ n = (e^{\bet \omk} -1)^{-1}~, \omk =\sqrt{m_{\pi}^2 +\vk^2} $.
Though $n^+(\omp)$ and $n(\omk)$ are zero for the medium we are interested 
in, we retain them at this stage for generality and symmetry.

With these expressions for the propagators, we may carry out both the $x^0$ 
and $\vec x$ integrations in Eq.(3.9) and find the imaginary part from 
the energy denominators,
\be
Im\Gm(q)_{11}=\pi \tanh (\bet (q_0 -\mu)/2)Im \Gm (q),
\ee
where
\be
Im \Gm(q)=
-\int\frac{d^3p}{(2\pi)^32\omp}\int\frac{d^3k}{(2\pi)^32\omk}
(\ps-m)\left\{(1-n^-+n)\delta (q-p-k)+(n^-+n)\delta (q-p+k)\right\}.
\ee
Here the first term corresponds to $\eta\rw\pi N$ and the second 
to $\eta + \pi \rw N$, giving rise respectively to the discontinuity across 
the unitary and the `short' cuts \cite{Weldon}. For $\vq=0$, these extend
in the $E$-plane over $E\geq m+m_\pi$ and $0\leq E \leq m-m_\pi$. Note the
opposite sign before $n^-$ in the two terms; it ensures $Im \Gm
\sim O(\bn)$ for small $\bn$.
 
The complete result for $Im \Gm_{11}$ has also another piece given by 
\[-\int\frac{d^3p}{(2\pi)^32\omp}\int\frac{d^3k}{(2\pi)^32\omk}
(\ps+m)\left\{(1-n^++n)\delta (q+p+k)+(n^++n)\delta (q+p-k)\right\}, \]
which is non-zero for $E<0$ only, but has no term proportional to $n^-$.
Clearly this observation does not depend on the structure of vertices in 
the loop diagrams. We thus have the general result that there is no 
spectral function to one loop for $E<0$ in a medium with nucleons only.

The three momentum integration in Eq.(3.13) can be carried out immediately 
to get
\be
Im\Gm (E) = \pm \frac{f(\om)}{E}, ~~~~~~ \om= (E^2 +m^2 -m^2_{\pi})/2E~,
\ee
where 
\[ f(\om) = \frac{\sqrt{\om^2-m^2}}{8\pi^2} (\gz \om -m) n^- (\om), \] 
The +(--) sign in front in Eq.(3.14) correspond to the unitary (short) cut. 
Inserting this result in Eq.(2.3) we get the spectral representation for $\Gm$,
\be
\Gm (E)= \int \frac{dE' f(E')}{E'(E'-E)},
\ee
where the integration is understood to run over the two cuts for $E>0$, with
appropriate sign for $f(E)$.

Let us illustrate a convenient way to treat the integration in Eq.(3.15).
We write it explicitly as
\be
\Gm (E)= \int_{m+m_\pi}^\infty \frac{dE'f(E')}{E'(E'-E)}
-\int_0^{m-m_\pi}\frac{dE'f(E')}{E'(E'-E)}\, .
\ee
The range of the second integral can be mapped on to that of the first by 
the inverse transformation $E' \rw (m^2-m_\pi^2)/E'$. Noting that $\om$ 
and hence $f(\om)$ is form invariant under this transformation, we have
\be
\Gm (E)=\int_{m+m_\pi}^\infty\frac{dE'}{E'}f(E')
\left(\frac{1}{E'-E}-\frac{1}{(m^2-m_\pi^2)/E' -E}\right) .
\ee
Once we are on the unitary cut, the kinematics is determined by the first
$\de$-function in Eq.(3.13). With $\vq=\vec 0$, it gives 
\be
E=\sqrt{m^2+p^2}+\sqrt{m_\pi^2+p^2},
\ee
so that $E$ becomes the total energy of the $\pi N$ system in its 
centre-of-mass frame. The 3-momentum can be solved to give
\be
p^2= \{E^2-(m+m_\pi)^2\}\{E^2-(m-m_\pi)^2\}/4E^2.
\ee
 
The distribution function $n^- (\om)=\th (\mu - \om)$ restricts the upper
limit of the integral to $E_F$ given by Eq.(3.18) with $p$ replaced by the
Fermi momentum $p_F$. Approximating $f(E)$ to leading order in momentum, 
the $\gm$-matrix structure in Eq.(3.17) factorizes and we finally get, 
\be
\Gm (E)=A(E) (1-\gz),
\ee
with
\be
A(E)=-\frac{m}{8\pi^2}\int_{m+m_\pi}^{E_F}\frac{dE'p'}{E'}\left(\frac{1}{E'-E}-
\frac{1}{(m^2-m_\pi^2)/E'-E}\right).
\ee

The spectral representation of the remaining diagrams can be obtained in the
same way. For the vertex diagrams (f) and (g), we have 
\be
\Pi(q)_{(f)+(g)}=2\lm^2 S(q)\Lm (q),
\ee
where the imaginary part of $\Lm (q)$ is related to $\Gm (q)$ by
\be
Im\Lm(q)_{11}=- \frac{3g_A}{4\F} (\qs+m)Im\Gm (q)_{11}.
\ee
We can again simplify such expressions by restricting to leading order in 
momentum. A further simplification results in the chiral limit $(m_\pi = 0)$, 
in which we shall evaluate the sum rules. With these approximations, the 
$\gm$-matrix structure in $\Lambda$ factorises as in $\Gm$, getting
\be
\Lm (E)=B(E) (1-\gz),
\ee
with
\be
B(E)=\int \frac{dE' g(E')}{E'-E}~, ~~~ g(E)=\frac{n^-(\om)}{8\pi^2}\frac{3g_A}{4\F}
\left( \frac{E^2-m^2}{2E}\right)^2 .
\ee
With the replacement (2.5), Eq.(3.22) becomes
\be
\Pi (E)_{(f)+(g)} =-\frac{\lm^2}{m} B(E) (1-\gz) .
\ee
 
Finally the self-energy diagram (h) is given by
\be
\Pi(q)_{(h)}=-\lm^2 S(q)\Sg (q)\ S(q),
\ee
where the imaginary part of $\Sg(q)$ is again related to that of 
$\Gm$ by
\be
Im\Sg (q)_{11}=\frac{3g_A^2}{4\F}(\qs+m)Im \Gm (q)_{11}(\qs+m).
\ee
It gives the integral representation
\be
\Sg (E)=C(E) \gz\, ,
\ee
with 
\be
C(E)=\int \frac{dE' h(E')}{E'-E},~~~ h(E)=\frac{n^-(\om)}{4\pi^2}
\frac{3g_A^2}{4\F}\left( \frac{E^2-m^2}{2E} \right)^3 .
\ee
Eq.(3.27) may now be expressed as 
\be
\Pi (E)_{(h)}=- \frac{\lm^2}{4m^2} C(E)\gz -\frac{\lm^2}{2} \left\{
\frac{1}{m(E-m)} +\frac{1}{(E-m)^2} \right\} C(E) (1+\gz) .
\ee
So far the evaluation of graphs has yielded terms contributing to either the
nucleon pole or the $\pi N$ continuum. Now the last two terms in Eq.(3.31)
appear as a {\it product} of the pole (simple and double) and the continuum 
integral. For interpretation  and later convenience, we write them as a 
{\it sum} of such terms, getting \footnote{It may be obtained by expanding 
$C(E)$ in a Taylor series with remainder about the nucleon pole. Alternatively, 
we may resolve the product of energy denominators into partial fractions.}
\be 
\Pi (E)_{(h)} = -\frac{\lm^2}{4m^2} C(E)\gz -\frac{\lm^2}{2} 
\left\{\frac{c_1/m +c_2}{E-m} + \frac{c_1}{(E-m)^2} +\frac{1}{m}\int
\frac{dE' h(E')}{(E'-m)(E'-E)} + \int \frac{dE'h(E')}{(E'-m)^2 (E'-E)}
\right\} (1+\gz),
\ee
where
\[ c_1=\int\frac{dE h(E)}{E-m} =\frac{3g_A^2 \bn}{16\F}\, ,~~~~
c_2=\int \frac{dE h(E)}{(E-m)^2} =0. \]
This completes our evaluation of the Feynman graphs in terms of their 
spectral integrals.

As a side result, we may now read off the modified nucleon pole parameters
in nuclear medium. The dispersion integrals appearing above in the evaluation 
of graphs are all finite at $E=m$, though some are non-analytic, being
proportional to fractional powers of $\bn$. Collecting the results for the
simple and double poles, we see that the vacuum pole given by Eq.(3.1) is 
modified to 
\[-\frac{\lm^{\star 2}}{E-m^\star} \,\frac{1}{2} (1+\gz),\]
where
\bea
\lm^{\star 2}& =&\lm^2 \left[ 1+\left\{ \frac{3}{4m}\left(C_S -C_T
+\frac{g_A^2}{4\F}\right) -2\zeta \right\}\bn \right], \nonumber \\
m^\star &=& m +\left\{ \frac{3}{4} (C_S - C_T) + \frac{3g_A^2}{16\F}
\right\}\bn
\eea
At normal nuclear density, it gives $m^\star - m = -322$ MeV, that is much too
large compared to the value obtained from the self-consistent, relativistic 
Hartree-Fock calculation, namely $ m^\star -m =-87$ MeV \cite{Brockmann}.
We postpone discussing this point to Sec.V.

\section{Sum Rules}
\setcounter{equation}{0}
\renewcommand{\theequation}{4.\arabic{equation}}

To write the QCD sum rules, we need the Borel transform of the amplitudes
with respect to a convenient variable. In the case of vacuum amplitudes, the
dispersion variable $q^2$ in the space-like region is the the natural choice
for it. With $\vq$ fixed, we might think of replacing it by $E^2$ with $E$ on
the imaginary axis in the $E$-plane. However, as we pointed out earlier, 
the loop graphs for nuclear medium do not generate any density-dependent
singularity at all for negative values of $E$, precluding any 
symmetry of the amplitude under $E\rw -E$. We thus choose $Q=-E >0$ as 
the variable with respect to which we take the Borel transform.
It is then given by the differential operator,
\[\bt= \lim_{Q\rw\infty,\, n\rw \infty} \frac{1}{(n-1)!} Q^n \left (
-\frac{d}{dQ} \right)^n\,, \]
such that $Q/n =M $ is held fixed, replacing $Q$ by the new variable $M$.
Thus
\[\bt\left \{\frac{1}{(Q+m)^k}\right\}=\frac{1}{(k-1)!} \frac{1}{M^k}
e^{-m/M} \,.\]

It is now simple to find the Borel transforms of the contributions of 
individual diagrams evaluated in the last Section. To leading order in $\bn$
we get,
\bea
& &\bt\{\Pi_{(b)}\}=\frac{3}{16} (C_S-C_T)\left( 1-\frac{2m}{M}\right) 
\frac{F}{Mm} (1+\gz)~, \nonumber \\
& &\bt\{\Pi_{(c)+(d)}\}=-\zeta \frac{F}{M} (1+\gz)~, \nonumber \\
& &\bt\{\Pi_{(e)}\}=-\frac{3}{32\F}\frac{F}{M^2} (1-\gz)~, \nonumber \\
& &\bt\{\Pi_{(f)+(g)}\}=\frac{3g_A}{32\F} \frac{F}{Mm} (1-\gz)~, \nonumber \\
& &\bt\{\Pi_{(h)}\}= 0~,
\eea
where we have introduced, for short, $F=\lm^2\, \bn\, e^{-m/M}$. The last
equation shows clearly and simply the importance of the continuum
contribution in the low energy region, that is neglected in the commonly
used spectral ansatz for the two-point function. The self-energy graph (h)
does contribute to the pole term shifting both the residue and the position
as given by Eq.(3.33). As the Borel transform of the full contribution of
this graph is zero, the continuum part must be equal and opposite to the
pole term for large enough $Q$.  

We now turn to operator expansion. We specialize to the rest frame of 
the medium $u^{\mu} =(1,0) $ and for external momentum $q_{\mu}=(-Q,0)$.
Then the Fourier transform of the operator coefficients in Eq. (2.15) gives 
\bea
\Pi (Q)&=& \frac{1}{4\pi^2} Q^2ln(-Q^2) ( \la \bu u\ra
+4\la u^{\dag} u\ra \gz ) \nonumber \\
&+&\frac{5}{6\pi^2} Qln(-Q^2) \la \Th^f \ra\gz  \nonumber \\
&+& \frac{2}{3Q} ( {\la \bu u\ra}^2\gz +2\la\bu u\ra \la u^{\dag}u\ra ).
\eea

The ensemble average of a local operator can be expanded in powers of
nucleon number density. Thus for any operator $R$ we have 
the familiar result to first order,
\[ \la R\ra= \la0|R|0\ra +\int \frac{d^3p}{(2\pi)^3 2p_0} \sum_{\al}
\la p,\al|R|p,\al\ra n^- (p_0)\, , \]
where $|p,\al\ra $ is a one nucleon state of momentum $p$ and spin and 
isospin denoted jointly by $\al$ and normalized as 
\[ \la p,\al | p',\al'\ra = (2\pi)^3 2 p_0 \de_{\al,\al'}\de^4 (p-p') .\]
If the constant, averaged matrix element is denoted by $\la p|R|p\ra$,
it becomes 
\[\la R\ra =\la 0 |R|0\ra + \frac{\la p|R|p\ra}{2m} \bn . \]

We now apply this formula to the different operators appearing in Eq.(4.2).
For the operator $\bu u$ it gives
\be
\la \bu u \ra = \la 0|\bu u |0\ra +\frac{\sg}{2\hm} \bn ,
\ee
where the quark condensate in vacuum has the value,
$\la 0|\bu u |0\ra = \la 0|{\bar d}d |0\ra =- (225 MeV )^3,$
and $\hm$ is the averaged quark mass $\hm = (m_u+m_d)/2 = 7 MeV$. 
The so-called $\sg$-term \footnote{The $\sg$-term in this Section 
cannot be confused with the self-energy $\sg$ in Sec.III .} is
$ \sg =\hm \la p| \bu u + \bd d|p\ra /(2m) =45 MeV $ \cite{Gasser3}.
The matrix element for the operator $u^{\dag} u$ is also obtained, if we
identify the quark current with the nucleon current as, 
$ \bu \gmd u +\bd \gmd d =3\bar{\psi} \gmd \psi \,.$ 
We then get
\be
\la u^{\dag} u \ra =\frac{3}{2} \bn .
\ee
Finally to obtain $\la \Th^f\ra$, we need the nucleon matrix element of the
energy-momentum tensor, 
\[ \la p|\Th_{\mu\nu}^f | p\ra = 2 A^f (p_\mu p_\nu -
\frac{1}{4} g_{\mu\nu} p^2 ),\]
where the constant $A^f$ is given by an integral over the nucleon structure
function in the deep inelastic scattering. At $Q^2 = 1 GeV^2$,  
it gives $A^f=.62$ \cite{Martin}. We thus get
\be
\la \Th^f \ra =\frac{3}{4} m A^f \bn .
\ee

Before we can take the Borel transform of Eq.(4.2), we have to pay attention 
to the singularities generated by the logarithms. Writing 
\be
ln(-Q^2)=lnQ+ln(-Q)=ln(-E)+ln E,
\ee 
we see that the first and the second terms give rise to the imaginary parts
for $E > 0$ and $E<0$ respectively. But we have no singularities for $E<0$
on the spectral side. So we must also remove the second term in Eq. (4.6).
Then noting that 
\[ \bt \{ Q\,lnQ\}=M,~~~~\bt\{Q^2 lnQ\} =-2M^2 ,\]
Eq.(4.2) immediately gives
\bea 
\bt\{\Pi (Q)\} &=& \left\{ -\frac{M^2}{\pi^2} \left( \frac{\sg}{4\hm} 
+3\gz \right) +\frac{5}{8\pi^2} A^q M m \gz \right. \nonumber \\    
 & &\left. + \frac{2}{3M} \la 0 |\bu u| 0\ra 
\left(3+f\frac{\sg}{\hm}\gz \right) \right\} \bn
\eea

The only other ingredient needed to write the sum rules is the continuum
contribution at high energy on the spectral side. Following the usual
procedure, we extract the imaginary part of the two-point function in this
region from the logarithms present in Eq. (4.2) for the operator product
expansion. The corresponding amplitudes are given by dispersion integrals 
starting from $W$, say. Being proportional to the operator matrix elements, 
these spectral contributions may conveniently be transferred to the 
operator side.

It remains to equate the coefficients of $(1\mp \gz)$ in Eqs. (4.1) and
(4.7) and include the continuum contribution from the high energy region 
to get the desired QCD sum rules,
\bea 
&&\frac{3\lm^2}{16\F}\frac{e^{-m/M}}{M^2} \left( g_A\frac{M}{m} -1\right) 
\nonumber \\
&&=-\frac{2}{3M} \la 0|\bu u | 0\ra \left( f\frac{\sg}{\hm}-3 \right)
+\frac{M^2}{\pi^2} \left( 3-\frac{\sg}{4\hm} \right) V_2 -\frac{5}{8\pi^2}
A^f MmV_1~,  
\eea
and
\bea
& &2\lm^2 e^{-m/M} \left\{ -\frac{\zeta}{M} +\frac{3}{8Mm} (C_S
-C_T)\left( 1-\frac{m}{M} \right) \right\} \nonumber \\
& & = \frac{2}{3M} \la 0| \bu u| 0 \ra \left( f\frac{\sg}{\hm} +3 \right)
-\frac{M^2}{\pi^2} \left( 3+\frac{\sg}{4\hm}\right) V_2 
+\frac{5}{8\pi^2} A^f Mm V_1~,
\eea
where 
\[V_1=1-\left( 1+\frac{W}{M} \right) e^{-m/M},~~~ 
V_2=1-\left( 1+\frac{W}{M} +\frac{W^2}{2M^2} \right) e^{-m/M}. \]

The two sum rules contain three unknown constants $\lm,~~\zeta$ and $f$, besides
$W$, the lower end of the high energy continuum. The coupling constant $\lm$
is, of course, known from the vacuum sum rules \cite{Ioffe1}, but we consider 
the possibility of its independent determination from our sum rules.

As usual we look for an interval in the Borel parameter $M$, where the
spectral and the operator sides of a sum rule overlap, leading to a plateau 
in the curve plotting the value of a `constant' as a function of $M$. We vary
$W$ and $f$ in the range $.8 \leq W \leq 1.8$  and $0 \leq f \leq 1$  
respectively. Thus for each set of values of $W$ and $f$ in this range, we
search for a reasonable plateau in the plots of $\lm^2$ and $\zeta$ against
$M$.

We find that, in general, the constancy of $\lm^2$ extends over a wider
region in $M$ than for $\zeta$. In this situation we first select values for
the pair $(W,~f)$ for which there is such a constancy in $\lm^2$. Next from
among this set, we choose those for which the curve for $\zeta$ also shows a
reasonable plateau. The different curves so selected are shown in Figs. 2
(a) and (b). Our results for the ranges of different constants are
\be
W= (1.6 - 1.8)~ GeV,~~~ f=0.2 - 0.4,~~~ \lm^2 =(1.5 - 2.7)\times 10^{-3} GeV^6,
~~~\zeta = (20 - 65)~ GeV^{-3} 
\ee
It is remarkable that our sum rule result for $\lm^2$ is so close to that
obtained from the vacuum sum rules \cite{Ioffe1}.

\begin{figure}
\centerline{\psfig{figure=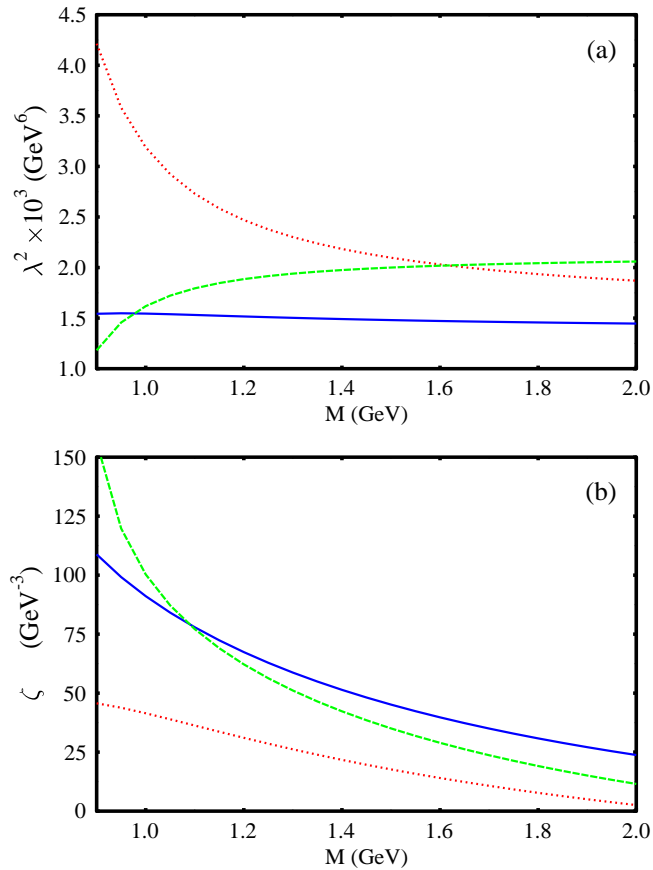,width=12cm,height=13cm}}
\caption{ Plot of $\lm^2$ and $\zeta$ as function of $M$. The solid, dotted
and the dashed graphs correspond to the values (1.6, 0.3),\, (1.6, 0.4),\,
and (1.8, 0.2) respectively for the set $(W, f)$.}
\end{figure}

\section{Discussion}

Instead of assuming an ansatz for the two-point function, as is usually 
done in the literature, we have calculated it in chiral perturbation theory. 
To one loop, the coupling constants appearing at the vertices in Feynman 
graphs are all known, except for $\zeta$ arising from the vertex 
$\eta \rw ({\bar \psi}\psi) \psi$. We determine it from the sum rules.

As we saw at the end of Sec.III, the four-nucleon couplings lead to too 
big a mass shift (self-energy) of the nucleon in normal nuclear matter. 
It is related to the fact that the formulation of chiral perturbation 
theory for the two-nucleon system faces a problem, due to the proximity 
of the virtual or bound states to the theshold of $N\!N$ scattering 
\cite{Weinberg}. It makes the effective coupling constants rather large.
Further, when multiplied by $\bn$ for normal nuclear matter, these are still
too large for the one-loop result to be valid \cite{Montano}. A resummation 
of the loop diagrams with these vertices is supposedly needed \cite{Kaplan}.

We must, however, point out that this breakdown of the one-loop result 
in normal nuclear matter does not prevent us from using this four-nucleon 
coupling for writing our sum rules. The point here
is that we are not working in normal nuclear matter. All we need is the
validity of an expansion of the two-point function in powers of $\bn$ for
sufficiently small $\bn$. Since any resummation can only alter coefficients 
of terms quadratic and higher in $\bn$, our use of the coefficient of $\bn$
to one loop remains valid.

Our calculation does not support an assumption in the popular ansatz for 
the two-point function, namely that it can be well-approximated by the 
(dressed) nucleon pole term and an integral over the {\it high} energy region. 
We find that each Feynman graph, besides contributing possibly to the pole, 
also contributes an integral over the {\it low} energy region, which is not 
accounted for by the ansatz. Also there are no density dependent discontinuities 
of the amplituide for negative energy in nuclear medium. Hence the variable to 
write the dispersion integrals is $E$ and not $E^2$. 

Finally we suggest a different use of the sum rules. There is a tendency in
the literature to try to predict the nucleon self-energy from the sum rules.
Indeed, this is in accordance with the works on the vacuum sum rules, where
these are used to find the masses and the couplings of the single particles 
communicating with the currents in the two-point functions. However, it
appears to us that the sum rules in the medium, as we have written,
may be utilised more profitably to find the condensates in the medium.

A way to carry out such a program would be to extend the sum rules to
non-zero external three-momenta $\vq$. Here we retain the different
four-quark operators as such, without invoking factorization. With the
additional variable $\vq \,^2$ at hand, it should be possible to extract
numerical values for these oprator condensates in nuclear medium. These
values may then be compared with model calculations \cite{Faessler}.   
 
\section*{Acknowledgments}

One of us (H. M.) wishes to thank Saha Institute for Nuclear Physics, India
for warm hospitality. The other (S.M.) acknowledges support of CSIR, 
Government of India.

\end{document}